\documentclass[%
 aapm,
 mph,%
 amsmath,amssymb,
 reprint,%
]{revtex4-2}

\usepackage{graphicx}
\usepackage{dcolumn}
\usepackage{bm}
\usepackage{textcomp}
\usepackage{float}

\begin{document}

\preprint{AAPM/123-QED}

\title{Dynamics and efficient conversion of excitons to trions in non-uniformly strained monolayer WS$_2$}

\author{Moshe G. Harats}\author{Jan N. Kirchhof}\author{Mengxiong Qiao}\author{Kyrylo Greben}\author{Kirill I. Bolotin}
 \affiliation{Department of Physics, Freie Universit\"{a}t Berlin, 14195 Berlin, Germany}
\email{moshe.harats@fu-berlin.de}

\begin{abstract}
We investigate the transport of excitons and trions in monolayer semiconductor WS$_2$ subjected to controlled non-uniform mechanical strain.  We actively control and tune the strain profiles with an AFM-based setup in which the monolayer is indented by an AFM tip. Optical spectroscopy is used to reveal the dynamics of the excited carriers.  The non-uniform strain configuration locally changes the valence and conduction bands of WS$_2$, giving rise to effective forces attracting excitons and trions towards the point of maximum strain underneath the AFM tip. We observe large changes in the photoluminescence spectra of WS$_2$ under strain, which we interpret using a drift-diffusion model. We show that the transport of neutral excitons, a process that was previously thought to be efficient in non-uniformly strained 2D semiconductors and termed as “funneling”, is negligible at room temperature in contrast to previous observations. Conversely, we discover that redistribution of free carriers under non-uniform strain profiles leads to highly efficient conversion of excitons to trions. Conversion efficiency reaches $\simeq 100\%$ even without electrical gating. Our results explain inconsistencies in previous experiments and pave the way towards new types of optoelectronic devices.
\end{abstract}

\maketitle

\section{Introduction}

Two-dimensional materials from the class of transition metal dichalcogenides (TMDCs) are actively considered for applications in photonics, electronics, and optoelectronics. TMDCs feature a direct band-gap in the monolayer limit \cite{Mak2010,Zhao2013}, exhibit an unusual spin/valley locking \cite{Xu2014,Xiao2012CoupledDichalcogenides,Mak2014},and can host tunable single photon emitters  \cite{Tonndorf2015,Chakraborty2015,Koperski2015,Srivastava2015,He2015}. Prototype TMDC-based electronic devices including transistors \cite{Roy2014Field-EffectComponents}, p-n junctions  \cite{Lee2014}, and solar photoconversion devices  \cite{Jariwala2017VanOutlook,Akama2017} have already been demonstrated. Moreover, high Young's modulus of TMDCs ($\sim 170 GPa$ in WS$_2$ \cite{Liu2014}) invites applications of these materials in flexible electronics  \cite{Lee2008,Liu2014,Roldan2015}.

Physical properties of TMDCs change under mechanical strain \cite{Feng2012}. In the simple case of constant uniaxial strain, the band-gap energy is reduced by $50\;meV/\%$ \cite{Conley2013,Niehues2018} and the phonon-assisted coupling is altered \cite{Christiansen2017,Niehues2018}. The band-gap reduction is twice higher, $100\;meV/\%$, for uniform biaxial strain \cite{Lloyd2016}. In conventional semiconductor materials, the advent of "strain-engineering", controlled strain-induced modification of the band-gap, allowed directing flows of excitons in quantum wells \cite{Lazic2014} and led to the performance improvement of strained Silicon MOSFET transistors \cite{Manasevit1982,People1984ModulationHeterostructures}. In the same vein, strain-engineering of TMDCs has recently been analyzed. The theoretical proposal of Ref. \onlinecite{Feng2012} considered a spatially-varying strain profile induced in a suspended TMDC membrane by a sharp tip of an atomic force microscope (AFM). In such a configuration, a force proportional to the band-gap energy gradient acts on a photoexcited  electron/hole pair (an exciton), transporting it to the center of the strain "funnel" at the location of the tip (Fig. \ref{Fig1}(b)). Two features make this setup attractive for efficient solar photoconversion. First, spatial variation of the local bandgap broadens the absorption spectrum of the TMDC. Second, excitons transported to the location of the AFM tip can be efficiently extracted and converted to electrical current. Experimental signatures consistent with the funnel effect have been previously observed in wrinkled few-layer MoS$_2$ \cite{Castellanos-Gomez2013} and monolayer MoS$_2$ nanobubbles \cite{Tyurnina2019}. Nevertheless, these previous experiments did not allow for induction of predictable strain profiles, dynamic control and tunability of strain magnitude, or quantitative analysis of the funneling efficiency. Because of that, the mechanisms governing transport and dynamics of excitons in non-uniform strain profile has not been fully investigated.

Here, we experimentally realized the setup originally proposed by Ref. \onlinecite{Feng2012}. Highly non-uniform and  \textit{in-situ} tunable strain profiles are induced in suspended monolayer WS$_2$ by a tip of an all-electrical AFM both in ambient and under vacuum conditions, while optical fingerprints of funneling are collected by a high-resolution optical system. We observe large changes in the PL spectra as a function of strain. By comparing our results to a simple drift-diffusion model, we decouple the contributions of two effects: funneling of excitons and trions (charged excitons) and funneling of free charge carriers. Contrary to prior expectation, we find that funneling of excitons is a very inefficient process with less than $4\%$ of photoexcited excitons reaching the funnel center even at highest achievable strain. In contrast, funneled free carriers are found to dominate optical spectra by binding to neutral excitons to form trions with a conversion efficiency up to $\sim100\%$. Taken together, our results explain inconsistencies in prior experiments and open a new pathway towards ultra-efficient TMDC-based photoconversion and optoelectronic devices.

\section{Results}
\begin{figure}
\centering
\includegraphics[width=0.45\textwidth]{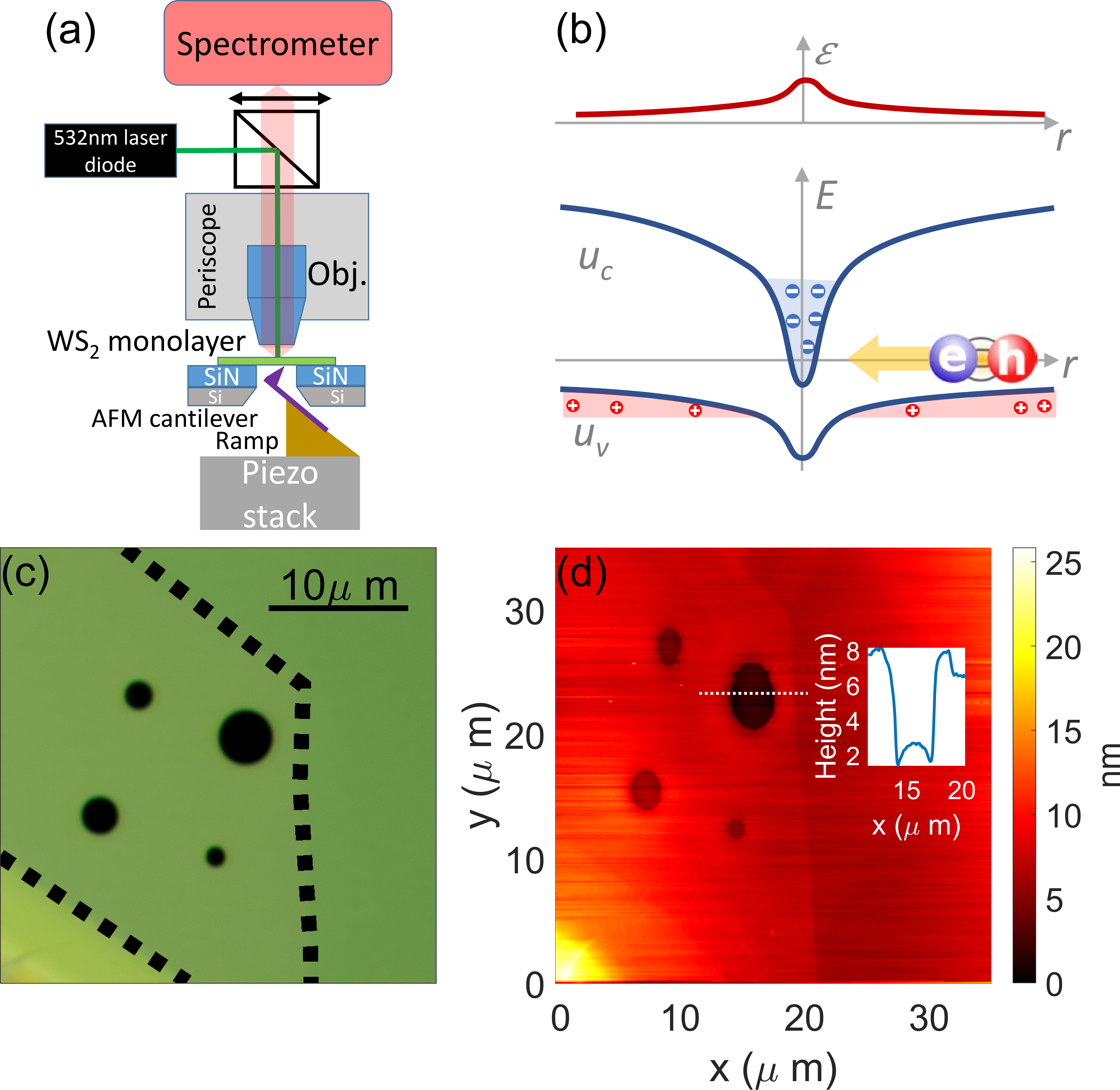}
\caption{(a) Schematics of the measurement apparatus. The WS$_2$ flake suspended on top of a hole in the SiN/Si membrane is indented from the bottom by an AFM cantilever while being optically interrogated from the top. (b) The schematic strain profile (top red curve) and band structure (bottom blue curves) of WS$_2$ vs. distance from the membrane center under non-uniform straining in our experiment. An effective force drives photoexcited excitons towards the point of minimum bandgap in the middle of the membrane where electrons are funneled and holes are inverse funnelled. (c) Optical image of a WS$_2$ monolayer (dashed black line) on top of perforated SiN/Si substrate.  (d) A tapping mode scan  of WS$_2$ shown in (c) recorded using our all-electrical AFM setup. The inset shows the height profile along a line crossing a suspended flake.}
\label{Fig1}
\end{figure}

To study a controllably strained TMDC, we require a suspended sample which can be approached from one side by an AFM tip while allowing optical access for excitation and detection from the other side. To produce suspended samples, we perforated holes with different diameters ranging from 0.5 to 2$\;\mu m$ in a $50\;nm$-thick Silicon Nitride (SiNx) using a Focused Ion Beam (FIB) milling. Monolayer WS$_2$ was mechanically exfoliated onto a PDMS film and  transferred on top of the holes using a dry-transfer technique \cite{Castellanos-Gomez2014DeterministicStamping} (Fig. \ref{Fig1}(c)). Full details on the fabrication of the samples is given in Ref.  \onlinecite{SeeOnline}.

To fulfill the main experimental challenge of this work, the induction of well-defined and controlled non-uniform strain profiles in a suspended monolayer of WS$_2$, we developed a novel AFM-based apparatus (Fig. \ref{Fig1}(a)).  A suspended monolayer is indented by an AFM tip from below, while being optically excited from above. Critically, the AFM cantilever is based on a piezo-resistive technology allowing all-electric motion actuation and deflection readout \cite{Dukic2015}. All-electrical operation of the cantilever allows for optical excitation/detection of WS$_2$ without the disturbing effect of laser sources normally employed in conventional AFMs. The described AFM setup is capable of topographic imaging both in tapping (see Fig. \ref{Fig1}(d)) and contact mode, recording force-distance curve for nanoidentation experiments \cite{Lee2014,Castellanos-Gomez2012,Zhang2016}, and, most importantly, applying a constant force on the suspended membrane whilst optical measurements are performed. Last but not least, the whole AFM system, being all-electrical and compact, can be incorporated into an optical cryostat, allowing additional measurements of the system under vacuum.

The optical part of the setup consists of a high-resolution objective (NA=$0.75$) and a periscope mounted on a scanning stage capable of nanometer resolution positioning. The sample is excited by a CW laser source at $532\;nm$  (spot size FWHM $600\;nm$, power $30\;\mu W$ unless specified otherwise) and the photoluminescence (PL) is directed to a spectrometer (Andor). Overall, our unique AFM/spectroscopy setup combines full AFM and full spectroscopic characterization capabilities. This combination is critical for the investigation of exciton transport in non-uniform strain profiles both in ambient and vacuum conditions.

We start by spatially locating a pristine WS$_2$ monolayer suspended over a hole. This is done using tapping mode AFM imaging to avoid puncturing delicate monolayers \cite{Tonndorf2015}. After locating the center of the suspended portion of the WS$_2$ monolayer, we perform a nanoidentation experiment at that location \cite{Castellanos-Gomez2012,Lee2008,Zhang2016}. From this data, we extract the Young's modulus ($Y$) and pre-tension ($\sigma_0$) of the membrane, the parameters necessary for the accurate determination of the strain profile \cite{Vella2017,SeeOnline}. Once preliminary characterization is complete, the sample is controllably indented by locking the PID loop at the desired force value, with the AFM tip still positioned at the center of the membrane (determined from maximal red-shift of the PL with respect to zero strain), and PL spectra vs. strain are acquired.

\begin{figure}[h]
\centering
\includegraphics[width=0.95\columnwidth]{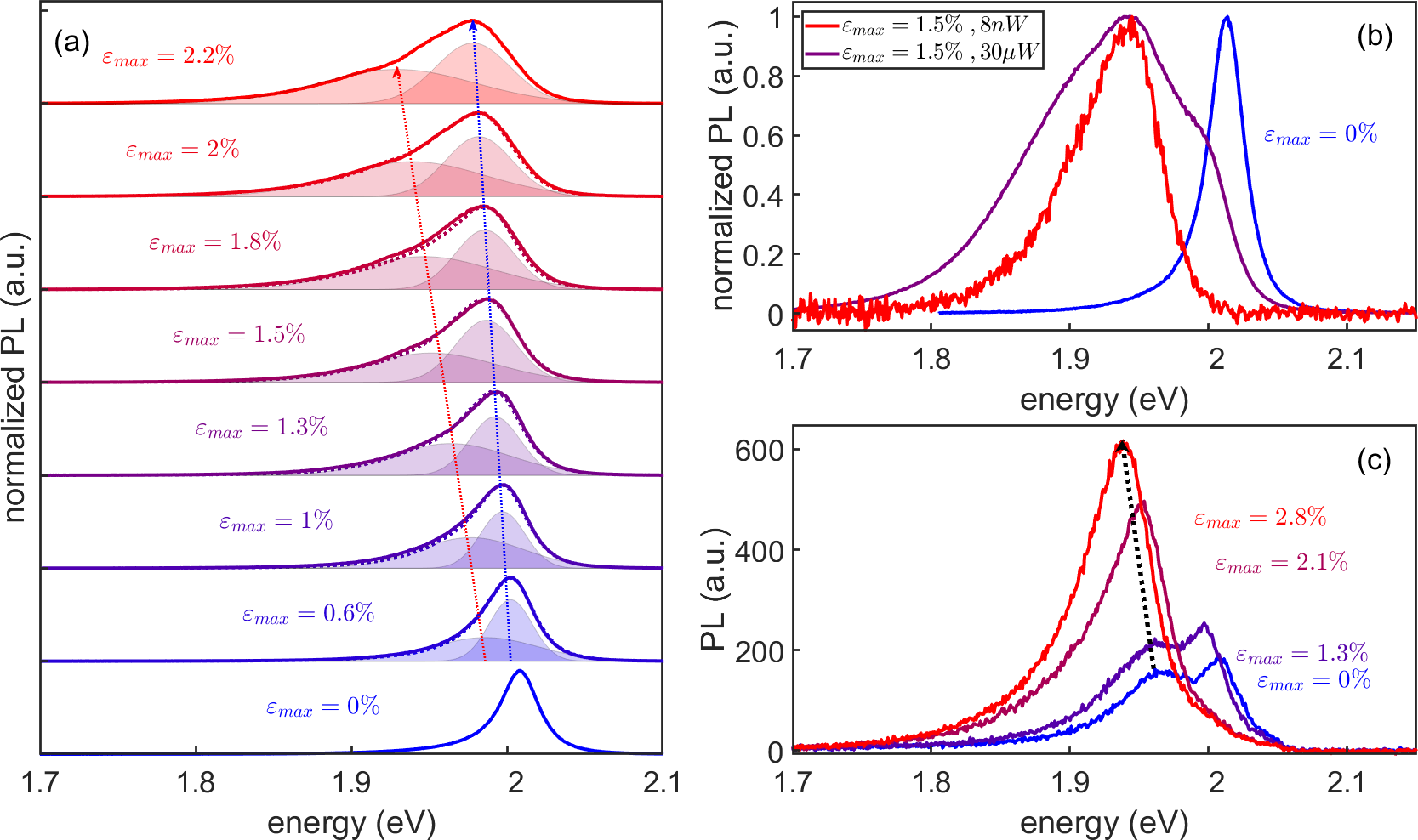}
\caption{(a) PL spectra of sample A at various strain levels - normalized and shifted for clarity. The data recorded during loading (solid lines) and unloading (dashed lines) cycles exhibit no hysteresis indicating the absence of mechanical slipping in our experiments. Maximal strain $\varepsilon_{max}$ (reached underneath the AFM tip) is shown next to each curve; The curve with  $\varepsilon_{max}=0\%$ corresponds to  an unstrained device. Two-peak Gaussian fits are shown along with the data. (b) Power- and strain- dependent PL spectra of sample B. The blue and purple curves are PL recorded at the excitation power of  $30\mu W$ for unstrained and strained ($\varepsilon_{max}=1.5\%$) device.  The red curve corresponds to the same strain, but with PL spectrum recorded at $8nW$. The data is normalized for better visibility.  (c) Strain-dependent PL spectra for sample C that was measured in vacuum. Well-resolved neutral and charged exciton peaks  evident at zero strain indicate high doping level in that device.}
\label{Fig2}
\end{figure}

Figure \ref{Fig2}(a) shows the evolution of the PL spectra of sample A as it is progressively indented.  Non-uniform strain profiles are parameterized by a single value, the maximum strain $\varepsilon_{max}$ that is reached at the middle of the membrane \cite{SeeOnline}. The sample is indented up to its breaking point, typically about $\varepsilon_{max} \sim 2.5\%$ (limited by the rupture of the flake caused by the sharp AFM tip). At zero strain, the PL spectrum is described by a non-symmetric Gaussian peak \cite{Niehues2018,Christiansen2017}. As strain is increased, this peak broadens and finally evolves at high strain into a two-peak structure: the red-shifted 'red' peak  and the blue-shifted  'blue' peak. Similar strain-dependent two-peak structure is seen in every measured sample, for example in sample B (Fig. \ref{Fig2}(b)). Amplitudes of the two peaks are sample-dependent. In sample B, for instance, the 'red' peak has much higher spectral weight compared to sample A.

Previous work \cite{Castellanos-Gomez2013} suggests a tempting interpretation of the two-peak structure. The 'red' peak could stem from emission of the excitons funneled to the point of the highest strain, whereas the 'blue' peak - to excitons that did not reach the funnel center, thus emitting throughout the sample as the laser excitation spot  exceeds the characteristics funneling length. A very high density of excitons is expected at the membrane's center in this interpretation. This should lead, in turn, to a rapid non-radiative Auger recombination of excitons which is known to be effective in TMDCs \cite{Kulig2018}. To test the role of Auger recombination, we recorded PL spectra for sample B at a relatively low power of $8nW$. At this power, only a few excitons are present in the entire sample at any given time and the role of Auger recombination should be negligible \cite{Kulig2018}. Figure \ref{Fig2}(b) shows that only the 'red' peak remains at low power while the 'blue' peak vanishes. In principle, such behavior is consistent with reduced Auger recombination and implies more efficient funneling at low power.

Finally, we tested for the contribution of charge excitons (trions) emission in our sample, as the binding energy of charged excitons in WS$_2$ is suspiciously close to the energy separation between the 'red' and 'blue' peaks. While PL spectra of samples A and B do not exhibit any trion contribution at zero strain, we later demonstrate that such contribution can arise when strain is increased. Therefore, we n-doped sample C by measuring it in vacuum. The desorption of water and Nitrogen from the sample surface increases the density of free electrons \cite{Ovchinnikov2014Electricalsub2/sub}, which, in turn, bind to neutral excitons to form negatively charged trions. Indeed, well-understood peaks corresponding to neutral (at $2.01\;eV$) and charged excitons (at $1.965\;eV$) are seen in sample C at zero strain (Fig. \ref{Fig2}(c)). Interestingly, as strain is increased, only the 'red' peak grows, while the 'blue' peak vanishes.

Overall, the experimental data of Fig. \ref{Fig2} poses the following questions. What is the physical origin of the 'red' and 'blue' peaks? How efficient is the funneling of neutral excitons in our sample? What is the role of charged excitons and why does their contribution appear to be strain-dependent?

\onecolumngrid

\twocolumngrid

\section{Discussion}\label{Theory}
To answer these questions, we analyze the drift-diffusion equations governing exciton transport in our system. For non-uniform density of excitons $n(r)$, the steady-state continuity condition for excitonic diffusion  $\vec{J_D}=D\nabla n(r)$ and  drift  $\vec{J_\mu}=\mu n(r)\nabla u(r)$ currents yields \cite{Kulig2018}:

\begin{multline}
\label{eq:diff_drift}
\nabla(D\nabla n(r))+\nabla(\mu\; n(r)\nabla u(r))- \\
\frac{n(r)}{\tau}-n^2(r)R_A+S(r)=0
\end{multline}

Here $D$ is the diffusion coefficient, $\mu=\frac{D}{k_BT}$ is the mobility, $R_A$ -- the Auger recombination rate, $\tau$ -- the exciton lifetime, and $S(r)=\frac{I_0}{2\pi \sigma^2} e^{-r^2/2\sigma^2}$ -- the exciton generation rate in a Gaussian illumination profile with intensity $I_0$ and $\sigma=FWHM/2 \sqrt{2\ln 2}$. Unless stated otherwise, we use material constants $D=0.3\;cm^2/s$, $\mu=12\;\frac{cm^2}{eV\cdot s}$ , $R_A=0.14\;cm^2/s$, and $\tau=1.1\;ns$ experimentally determined in Ref.  \onlinecite{Kulig2018}.  The change of the band-gap due to the strain is assumed to be  $u(r)=E_g-0.05\cdot\varepsilon(r)$, where $\varepsilon(r)$ is the trace of the strain tensor \cite{Conley2013,Lloyd2016}. A detailed analysis and simulations of Eq. \ref{eq:diff_drift} along with various calculations of $u(r)$ are shown in \cite{SeeOnline}.

\begin{figure}
\centering
\includegraphics[width=0.95\columnwidth]{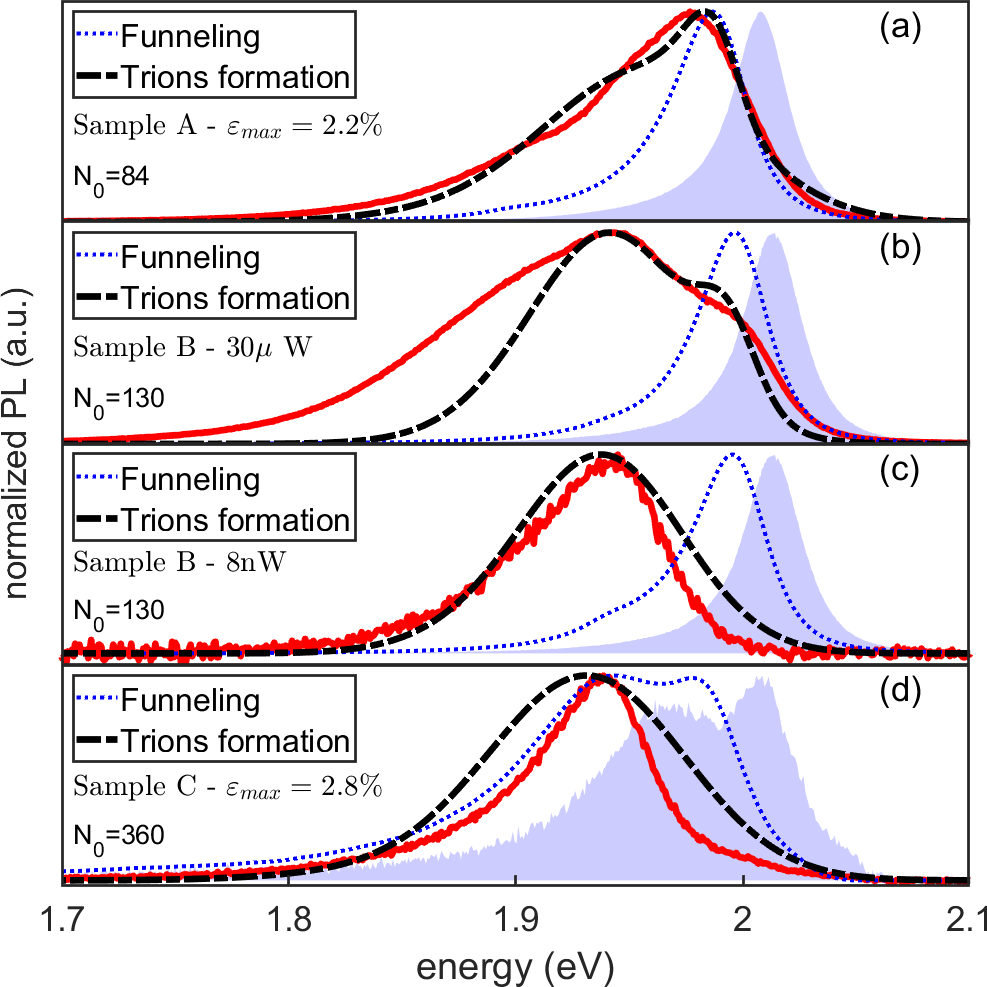}
\caption{
Comparison between the experimentally measured PL spectra in all measured samples (red curves) and the predictions of two models considered in the text: the model of Eq.  \ref{eq:diff_drift}  which only includes funneling of neutral excitons (thin dashed blue line), and the full model of Eq. \ref{eq:over_int} that adds the effects of carrier funneling and neutral-to-charged exciton conversion (thick dashed black line). The shaded blue area is the unstrained PL shown for reference. The spectra from the following devices are shown: (a) Sample A for the highest strain. (b)-(c) Sample B for high (b) and low (c) laser excitation intensities and the highest strain. (d) Sample C for the highest strain.}
\label{Fig3}
\end{figure}

Figure \ref{Fig3} (blue dashed curves) shows the PL spectra obtained from the numerical solutions of Eq. \ref{eq:diff_drift} \cite{SeeOnline}. These solutions clearly do not match the  experimental data (red curves). Indeed, while the 'red' peak of Fig. 2 could be interpreted as corresponding to a very efficient funneling process, the numerical solution of Eq. \ref{eq:diff_drift} exhibit funneling efficiency (defined as the fraction of all photoexcited excitons reaching the location of the AFM tip) that never exceeds $4\%$ \cite{SeeOnline}.

It is instructive to develop intuitive understanding for the observed low funneling efficiency. While the drift term in the Eq. \ref{eq:diff_drift} 'pushes' the exciton towards the funnel center with the force proportional to $\nabla u(r)$, the diffusion term randomizes that motion (see Fig. \ref{Fig1}(b)).  The average distances travelled by an exciton during its lifetime due to drift and diffusion respectively can be evaluated within a simple Drude approximation. We find that the diffusion length $l_{diffusion}=\sqrt{D\tau}\simeq 180nm$ is much larger compared to the drift length averaged over the excitation spot,  $<l_{drift}>=<\nabla u(r)>\mu\tau\simeq 5nm$. Dominating contribution of the diffusion leads, in turn, to inefficient funneling. One could claim that we observe a rather low funneling efficiency due to charging in the system \cite{Feng2012}. This is not the case here as our analysis using the drift-diffusion equation is blind to any charging effects and is also valid to a type I funnel that does not exhibit charging effects.

We can analyze the relative contributions of drift and diffusion in another, more quantitative way. It is easy to show that, on average, exciton current flows towards the funnel center (drift dominates over diffusion) if the following condition is met \cite{SeeOnline}:
\begin{equation}\label{eq:d_over_mu}
k_BT<-\frac{S(r)\nabla u(r)}{\nabla S(r)}
\end{equation}

\begin{figure}
\centering
\includegraphics[width=0.95\columnwidth]{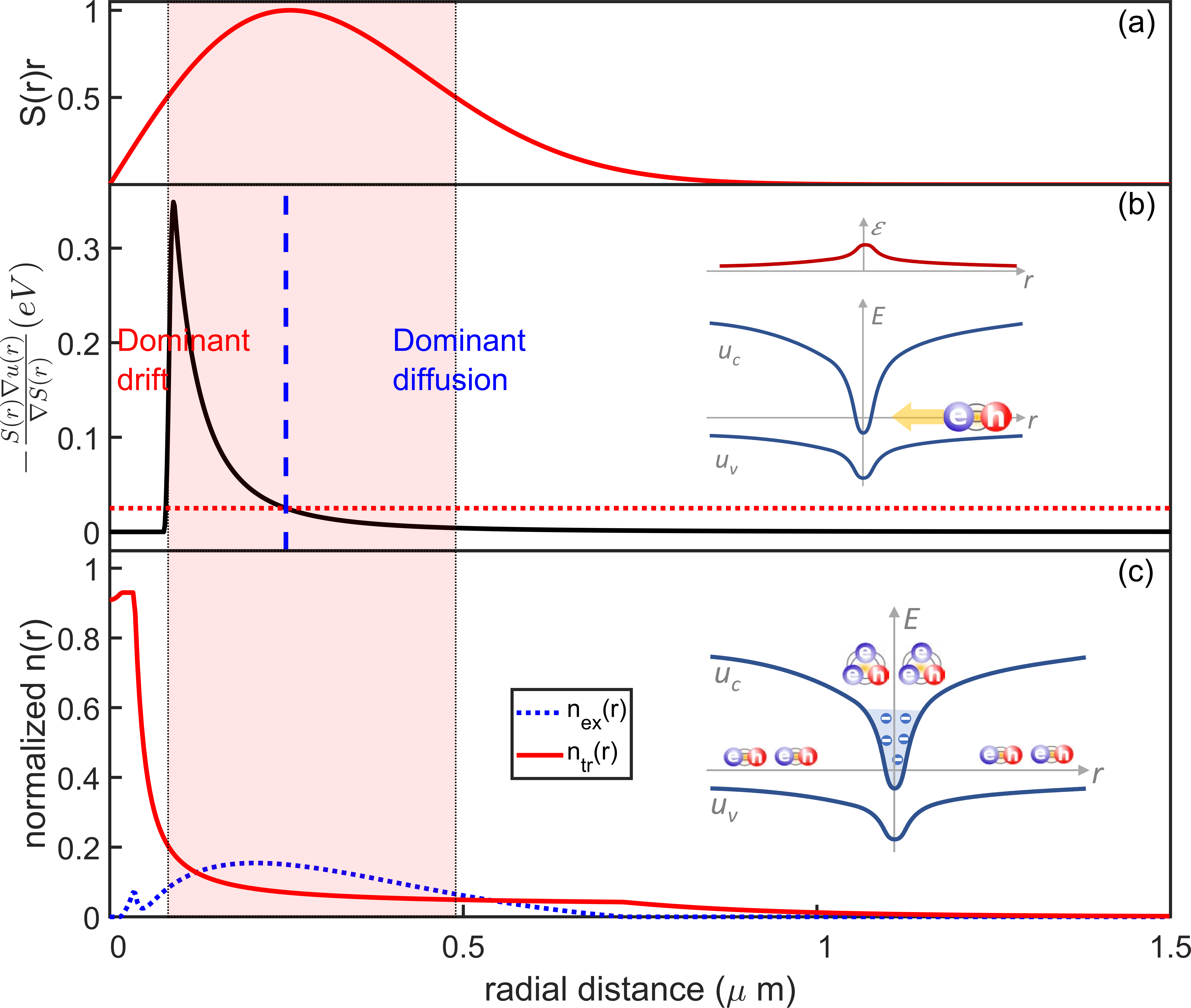}
\caption{(a) The source term $S(r)\cdot r$ in Eq. \ref{eq:diff_drift} corresponding to the illumination spot in our experiments. The shaded region, defined as the region where the term falls to less than 1/2 from the maximal value, represents the sample area producing the dominant contribution to the measured PL. (b) The ratio between the drift and the diffusion terms in the Eq. \ref{eq:diff_drift} (solid black curve). Equation \ref{eq:d_over_mu}  indicates that when this ratio is higher than $k_BT=25\;meV$ (dashed red line), drift dominates over diffusion. Inset:  A sketch of the forces acting on a neutral exciton in our straining conditions. (c) Normalized densities of neutral $n_{ex}(r)$ (dotted blue curve), and charged excitons $n_{tr}(r)$ (solid red curve) calculated using the formalism described in the text.  Large $n_{tr}(r)$ near the center of the membrane reflects effective funneling of free electrons towards the point of the highest strain, followed by their binding into trions. Inset: A sketch of the carrier funneling and trion conversion processes.}
\label{Fig4}
\end{figure}

Both the left-hand side  and the right-hand side of this formula are plotted in Figure \ref{Fig4}(b) as red dotted line and as black solid line respectively. We see that the condition above is only fulfilled for the small portion of the membrane ($r<250nm$), and that diffusion term dominates the rest of the membrane leading to inefficient funneling. We therefore posit that funneling cannot be as efficient at room temperature as predicted in Ref. \onlinecite{Feng2012}. We note that Eq. 2 also suggests that higher funneling efficiency may be possible at cryogenic temperatures.

If funneling is so inefficient, what other physical mechanism is responsible for the data of Fig. \ref{Fig2}? The hint comes from the data of sample C suggesting that the 'red' peak at high strain evolves from the trion peak at zero strain. To include the contribution of trions into our model, we again use the same Eq. \ref{eq:diff_drift}, but with $n(r)=n_{ex}(r)+n_{tr}(r)$ where $n_{ex}(n_{tr})$ is the exciton (negatively charged trion) density, respectively. Although there are differences in the physical constants $D,\mu, R_A, \tau$ between excitons and trions, we have found that the solution of $n(r)$ does not change significantly in a broad range of possible values \cite{SeeOnline}, thus we used the same values for both species.

It is easy to see that the density of trions near the device center is expected to be strongly strain-dependent. Indeed, the relative densities of neutral and charged excitons depend on the density of background electrons (doping level) $n_b(r)$ in our device. While at zero strain background carriers are uniformly distributed throughout the device, the applied nonuniform strain lowers the top of the conduction band $u_c(r)$. Quantitatively, assuming that the density of background electrons is described by the Boltzmann distribution, we obtain  $n_B(r)=\frac{N_0e^{ \Delta u_{c}(r)/k_BT}}{\int e^{ \Delta u_{c}(r)/k_BT}rdr}$ \cite{SeeOnline}. Here  $N_0$, which is strain dependent, represents the number of free carriers in the whole area of the strained membrane for any given time, and $\Delta u_{c}(r)$ is change of the energy of the top of the conduction band from the zero strain value \cite{SeeOnline}. The expression above makes it clear that the electrons are effectively 'funneled' towards the point of the highest strain at the center of the membrane.  As a consequence, photoexcited neutral excitons present near the membrane center bind to free electrons forming trions. To quantitatively determine the intensity of trion emission, Eq. \ref{eq:diff_drift} is solved for $n(r)$ as before, and $n_{ex},n_{tr}$ are determined from $n_B(r)$ using the law of mass action \cite{Siviniant1999ChemicalWells,Ross2013ElectricalSemiconductor,SeeOnline}.

Once the carriers densities $n_{ex},n_{tr}$ are determined, the entire PL spectrum is calculated by the following expression:

\begin{multline}\label{eq:over_int}
\left< PL \right> =\int_0^\infty  [ PL_{ex}(u_{ex}(r))n_{ex}(r)+ \\
PL_{tr}(u_{tr}(r))n_{tr}(r)] rdr
\end{multline}

Here $PL_{ex}(PL_{tr})$ are the spectral lines of the excitons (trions), respectively and are taken from the Gaussian fits of the spectrum at zero strain. Figure \ref{Fig3} shows a comparison between the model (black dashed line), using $N_0$ as a single fit parameter, and the experimental results (red line) for relevant strain and excitation intensities for all samples. The model is in much better agreement with the experimental data, especially in comparison with the model that does not include trion effects (thin blue dashed line). We therefore conclude that strain-related free carrier funneling, followed by conversion of neutral to negatively charged excitons, is the dominant process in our samples.

To  illustrate the mechanism responsible for the appearance of the strain-dependent trion contribution,  we show in Fig. \ref{Fig4}(c) the calculated spatial dependencies of neutral and charge exciton densities in Sample A (calculations for other samples are shown in Ref.  \onlinecite{SeeOnline}). We see that the trion density $n_{tr}(r)$ steadily increases and becomes much larger than $n_{ex}(r)$ towards the center of the funnel, consistent with funneling of free carriers to that region. At the same time, in the region of the sample that predominantly contributes to the observed signal due to the Jacobian $\textbf{r}dr$ (defined as the shaded area in Fig. \ref{Fig4}(a)), $n_{ex}(r)>n_{tr}(r)$. This explains comparable magnitudes of the trion and exciton peaks in Sample A (Fig. \ref{Fig3}(a). In contrast, in samples B (low excitation) and C the doping level is higher. In that situation, we find \cite{SeeOnline} that $n_{ex}(r)\ll n_{tr}(r)$ in the relevant area of the device, meaning that photoexcited neutral excitons are converted into trions with conversion efficiency approaching $100\%$.

Our findings suggest several important implications. First,  strain-dependent exciton-to-trion conversion produces experimental signatures that may appear similar to that of neutral excitons funneling, but much stronger in amplitude. This suggests that previous reports of exciton funneling might have strongly overestimated its efficiency. Second, while we experimentally realize, for the first time, the controlled funneling geometry of the theoretical proposal \cite{Feng2012}, the dominant process in such a device is found to be diffusion rather than drift, at least at room temperature. This means that the photoconversion mechanism proposed by Ref. \onlinecite{Feng2012} may not be feasible. Finally, and perhaps most importantly, the strain-dependent exciton-to-trion conversion may constitute another, more efficient photoconversion mechanism compared to that of Ref.  \onlinecite{Feng2012}. We speculate that the energetics of energy-harvesting of weakly-bound trions may be advantageous to that of strongly-bound neutral excitons.

To summarize, in this work we have presented a novel experimental setup that allows full dynamical control of strain amplitude and profile in optically-interrogated TMDC monolayers. We revealed that even in TMDCs strained to the point of breakage, the funneling of the excited carriers is not nearly as efficient as previously thought. On the other hand, we discovered that in the presence of non-uniform strain, another process, neutral-to-charged exciton conversion becomes dominant. It is noteworthy that this former process, while being physically very different than the latter, can produce similar experimental signatures leading to possible misinterpretations. Finally, we note that in the future it will be especially interesting to study the role of funneling at cryogenic temperatures, where the role of diffusion is minimized.

\bibliographystyle{aapmrev4-2}
\bibliography{Main}

\begin{thebibliography}{37}%
\makeatletter
\providecommand \@ifxundefined [1]{%
 \@ifx{#1\undefined}
}%
\providecommand \@ifnum [1]{%
 \ifnum #1\expandafter \@firstoftwo
 \else \expandafter \@secondoftwo
 \fi
}%
\providecommand \@ifx [1]{%
 \ifx #1\expandafter \@firstoftwo
 \else \expandafter \@secondoftwo
 \fi
}%
\providecommand \natexlab [1]{#1}%
\providecommand \enquote  [1]{``#1''}%
\providecommand \bibnamefont  [1]{#1}%
\providecommand \bibfnamefont [1]{#1}%
\providecommand \citenamefont [1]{#1}%
\providecommand \href@noop [0]{\@secondoftwo}%
\providecommand \href [0]{\begingroup \@sanitize@url \@href}%
\providecommand \@href[1]{\@@startlink{#1}\@@href}%
\providecommand \@@href[1]{\endgroup#1\@@endlink}%
\providecommand \@sanitize@url [0]{\catcode `\\12\catcode `\$12\catcode
  `\&12\catcode `\#12\catcode `\^12\catcode `\_12\catcode `\%12\relax}%
\providecommand \@@startlink[1]{}%
\providecommand \@@endlink[0]{}%
\providecommand \url  [0]{\begingroup\@sanitize@url \@url }%
\providecommand \@url [1]{\endgroup\@href {#1}{\urlprefix }}%
\providecommand \urlprefix  [0]{URL }%
\providecommand \Eprint [0]{\href }%
\providecommand \doibase [0]{https://doi.org/}%
\providecommand \selectlanguage [0]{\@gobble}%
\providecommand \bibinfo  [0]{\@secondoftwo}%
\providecommand \bibfield  [0]{\@secondoftwo}%
\providecommand \translation [1]{[#1]}%
\providecommand \BibitemOpen [0]{}%
\providecommand \bibitemStop [0]{}%
\providecommand \bibitemNoStop [0]{.\EOS\space}%
\providecommand \EOS [0]{\spacefactor3000\relax}%
\providecommand \BibitemShut  [1]{\csname bibitem#1\endcsname}%
\let\auto@bib@innerbib\@empty
\bibitem [{\citenamefont {Mak}\ \emph {et~al.}(2010)\citenamefont {Mak},
  \citenamefont {Lee}, \citenamefont {Hone}, \citenamefont {Shan},\ and\
  \citenamefont {Heinz}}]{Mak2010}%
  \BibitemOpen
  \bibfield  {author} {\bibinfo {author} {\bibfnamefont {K.~F.}\ \bibnamefont
  {Mak}}, \bibinfo {author} {\bibfnamefont {C.}~\bibnamefont {Lee}}, \bibinfo
  {author} {\bibfnamefont {J.}~\bibnamefont {Hone}}, \bibinfo {author}
  {\bibfnamefont {J.}~\bibnamefont {Shan}},\ and\ \bibinfo {author}
  {\bibfnamefont {T.~F.}\ \bibnamefont {Heinz}},\ }\bibfield  {title} {\enquote
  {\bibinfo {title} {{Atomically Thin MoS 2 : A New Direct-Gap
  Semiconductor}},}\ }\href {https://doi.org/10.1103/PhysRevLett.105.136805}
  {\bibfield  {journal} {\bibinfo  {journal} {Physical Review Letters}\
  }\textbf {\bibinfo {volume} {105}},\ \bibinfo {pages} {136805} (\bibinfo
  {year} {2010})}\BibitemShut {NoStop}%
\bibitem [{\citenamefont {Zhao}\ \emph {et~al.}(2013)\citenamefont {Zhao},
  \citenamefont {Ghorannevis}, \citenamefont {Chu}, \citenamefont {Toh},
  \citenamefont {Kloc}, \citenamefont {Tan},\ and\ \citenamefont
  {Eda}}]{Zhao2013}%
  \BibitemOpen
  \bibfield  {author} {\bibinfo {author} {\bibfnamefont {W.}~\bibnamefont
  {Zhao}}, \bibinfo {author} {\bibfnamefont {Z.}~\bibnamefont {Ghorannevis}},
  \bibinfo {author} {\bibfnamefont {L.}~\bibnamefont {Chu}}, \bibinfo {author}
  {\bibfnamefont {M.}~\bibnamefont {Toh}}, \bibinfo {author} {\bibfnamefont
  {C.}~\bibnamefont {Kloc}}, \bibinfo {author} {\bibfnamefont {P.-H.}\
  \bibnamefont {Tan}},\ and\ \bibinfo {author} {\bibfnamefont {G.}~\bibnamefont
  {Eda}},\ }\bibfield  {title} {\enquote {\bibinfo {title} {{Evolution of
  Electronic Structure in Atomically Thin Sheets of WS 2 and WSe 2}},}\ }\href
  {https://doi.org/10.1021/nn305275h} {\bibfield  {journal} {\bibinfo
  {journal} {ACS Nano}\ }\textbf {\bibinfo {volume} {7}},\ \bibinfo {pages}
  {791--797} (\bibinfo {year} {2013})}\BibitemShut {NoStop}%
\bibitem [{\citenamefont {Xu}\ \emph {et~al.}(2014)\citenamefont {Xu},
  \citenamefont {Yao}, \citenamefont {Xiao},\ and\ \citenamefont
  {Heinz}}]{Xu2014}%
  \BibitemOpen
  \bibfield  {author} {\bibinfo {author} {\bibfnamefont {X.}~\bibnamefont
  {Xu}}, \bibinfo {author} {\bibfnamefont {W.}~\bibnamefont {Yao}}, \bibinfo
  {author} {\bibfnamefont {D.}~\bibnamefont {Xiao}},\ and\ \bibinfo {author}
  {\bibfnamefont {T.~F.}\ \bibnamefont {Heinz}},\ }\bibfield  {title} {\enquote
  {\bibinfo {title} {{Spin and pseudospins in layered transition metal
  dichalcogenides}},}\ }\href {https://doi.org/10.1038/nphys2942} {\bibfield
  {journal} {\bibinfo  {journal} {Nature Physics}\ }\textbf {\bibinfo {volume}
  {10}},\ \bibinfo {pages} {343--350} (\bibinfo {year} {2014})}\BibitemShut
  {NoStop}%
\bibitem [{\citenamefont {Xiao}\ \emph {et~al.}(2012)\citenamefont {Xiao},
  \citenamefont {Liu}, \citenamefont {Feng}, \citenamefont {Xu},\ and\
  \citenamefont {Yao}}]{Xiao2012CoupledDichalcogenides}%
  \BibitemOpen
  \bibfield  {author} {\bibinfo {author} {\bibfnamefont {D.}~\bibnamefont
  {Xiao}}, \bibinfo {author} {\bibfnamefont {G.-B.}\ \bibnamefont {Liu}},
  \bibinfo {author} {\bibfnamefont {W.}~\bibnamefont {Feng}}, \bibinfo {author}
  {\bibfnamefont {X.}~\bibnamefont {Xu}},\ and\ \bibinfo {author}
  {\bibfnamefont {W.}~\bibnamefont {Yao}},\ }\bibfield  {title} {\enquote
  {\bibinfo {title} {{Coupled Spin and Valley Physics in Monolayers of MoS 2
  and Other Group-VI Dichalcogenides}},}\ }\href
  {https://doi.org/10.1103/PhysRevLett.108.196802} {\bibfield  {journal}
  {\bibinfo  {journal} {Physical Review Letters}\ }\textbf {\bibinfo {volume}
  {108}},\ \bibinfo {pages} {196802} (\bibinfo {year} {2012})}\BibitemShut
  {NoStop}%
\bibitem [{\citenamefont {Mak}\ \emph {et~al.}(2014)\citenamefont {Mak},
  \citenamefont {McGill}, \citenamefont {Park},\ and\ \citenamefont
  {McEuen}}]{Mak2014}%
  \BibitemOpen
  \bibfield  {author} {\bibinfo {author} {\bibfnamefont {K.~F.}\ \bibnamefont
  {Mak}}, \bibinfo {author} {\bibfnamefont {K.~L.}\ \bibnamefont {McGill}},
  \bibinfo {author} {\bibfnamefont {J.}~\bibnamefont {Park}},\ and\ \bibinfo
  {author} {\bibfnamefont {P.~L.}\ \bibnamefont {McEuen}},\ }\bibfield  {title}
  {\enquote {\bibinfo {title} {{Valleytronics. The valley Hall effect in MoS₂
  transistors.}}}\ }\href {https://doi.org/10.1126/science.1250140} {\bibfield
  {journal} {\bibinfo  {journal} {Science (New York, N.Y.)}\ }\textbf {\bibinfo
  {volume} {344}},\ \bibinfo {pages} {1489--92} (\bibinfo {year}
  {2014})}\BibitemShut {NoStop}%
\bibitem [{\citenamefont {Tonndorf}\ \emph {et~al.}(2015)\citenamefont
  {Tonndorf}, \citenamefont {Schmidt}, \citenamefont {Schneider}, \citenamefont
  {Kern}, \citenamefont {Buscema}, \citenamefont {Steele}, \citenamefont
  {Castellanos-Gomez}, \citenamefont {van~der Zant}, \citenamefont
  {Michaelis~de Vasconcellos},\ and\ \citenamefont
  {Bratschitsch}}]{Tonndorf2015}%
  \BibitemOpen
  \bibfield  {author} {\bibinfo {author} {\bibfnamefont {P.}~\bibnamefont
  {Tonndorf}}, \bibinfo {author} {\bibfnamefont {R.}~\bibnamefont {Schmidt}},
  \bibinfo {author} {\bibfnamefont {R.}~\bibnamefont {Schneider}}, \bibinfo
  {author} {\bibfnamefont {J.}~\bibnamefont {Kern}}, \bibinfo {author}
  {\bibfnamefont {M.}~\bibnamefont {Buscema}}, \bibinfo {author} {\bibfnamefont
  {G.~A.}\ \bibnamefont {Steele}}, \bibinfo {author} {\bibfnamefont
  {A.}~\bibnamefont {Castellanos-Gomez}}, \bibinfo {author} {\bibfnamefont
  {H.~S.~J.}\ \bibnamefont {van~der Zant}}, \bibinfo {author} {\bibfnamefont
  {S.}~\bibnamefont {Michaelis~de Vasconcellos}},\ and\ \bibinfo {author}
  {\bibfnamefont {R.}~\bibnamefont {Bratschitsch}},\ }\bibfield  {title}
  {\enquote {\bibinfo {title} {{Single-photon emission from localized excitons
  in an atomically thin semiconductor}},}\ }\href
  {https://doi.org/10.1364/OPTICA.2.000347} {\bibfield  {journal} {\bibinfo
  {journal} {Optica}\ }\textbf {\bibinfo {volume} {2}},\ \bibinfo {pages} {347}
  (\bibinfo {year} {2015})}\BibitemShut {NoStop}%
\bibitem [{\citenamefont {Chakraborty}\ \emph {et~al.}(2015)\citenamefont
  {Chakraborty}, \citenamefont {Kinnischtzke}, \citenamefont {Goodfellow},
  \citenamefont {Beams},\ and\ \citenamefont {Vamivakas}}]{Chakraborty2015}%
  \BibitemOpen
  \bibfield  {author} {\bibinfo {author} {\bibfnamefont {C.}~\bibnamefont
  {Chakraborty}}, \bibinfo {author} {\bibfnamefont {L.}~\bibnamefont
  {Kinnischtzke}}, \bibinfo {author} {\bibfnamefont {K.~M.}\ \bibnamefont
  {Goodfellow}}, \bibinfo {author} {\bibfnamefont {R.}~\bibnamefont {Beams}},\
  and\ \bibinfo {author} {\bibfnamefont {A.~N.}\ \bibnamefont {Vamivakas}},\
  }\bibfield  {title} {\enquote {\bibinfo {title} {{Voltage-controlled quantum
  light from an atomically thin semiconductor}},}\ }\href
  {https://doi.org/10.1038/nnano.2015.79} {\bibfield  {journal} {\bibinfo
  {journal} {Nature Nanotechnology}\ }\textbf {\bibinfo {volume} {10}},\
  \bibinfo {pages} {507--511} (\bibinfo {year} {2015})}\BibitemShut {NoStop}%
\bibitem [{\citenamefont {Koperski}\ \emph {et~al.}(2015)\citenamefont
  {Koperski}, \citenamefont {Nogajewski}, \citenamefont {Arora}, \citenamefont
  {Cherkez}, \citenamefont {Mallet}, \citenamefont {Veuillen}, \citenamefont
  {Marcus}, \citenamefont {Kossacki},\ and\ \citenamefont
  {Potemski}}]{Koperski2015}%
  \BibitemOpen
  \bibfield  {author} {\bibinfo {author} {\bibfnamefont {M.}~\bibnamefont
  {Koperski}}, \bibinfo {author} {\bibfnamefont {K.}~\bibnamefont
  {Nogajewski}}, \bibinfo {author} {\bibfnamefont {A.}~\bibnamefont {Arora}},
  \bibinfo {author} {\bibfnamefont {V.}~\bibnamefont {Cherkez}}, \bibinfo
  {author} {\bibfnamefont {P.}~\bibnamefont {Mallet}}, \bibinfo {author}
  {\bibfnamefont {J.-Y.}\ \bibnamefont {Veuillen}}, \bibinfo {author}
  {\bibfnamefont {J.}~\bibnamefont {Marcus}}, \bibinfo {author} {\bibfnamefont
  {P.}~\bibnamefont {Kossacki}},\ and\ \bibinfo {author} {\bibfnamefont
  {M.}~\bibnamefont {Potemski}},\ }\bibfield  {title} {\enquote {\bibinfo
  {title} {{Single photon emitters in exfoliated WSe2 structures}},}\ }\href
  {https://doi.org/10.1038/nnano.2015.67} {\bibfield  {journal} {\bibinfo
  {journal} {Nature Nanotechnology}\ }\textbf {\bibinfo {volume} {10}},\
  \bibinfo {pages} {503--506} (\bibinfo {year} {2015})}\BibitemShut {NoStop}%
\bibitem [{\citenamefont {Srivastava}\ \emph {et~al.}(2015)\citenamefont
  {Srivastava}, \citenamefont {Sidler}, \citenamefont {Allain}, \citenamefont
  {Lembke}, \citenamefont {Kis},\ and\ \citenamefont
  {Imamo{\u{g}}lu}}]{Srivastava2015}%
  \BibitemOpen
  \bibfield  {author} {\bibinfo {author} {\bibfnamefont {A.}~\bibnamefont
  {Srivastava}}, \bibinfo {author} {\bibfnamefont {M.}~\bibnamefont {Sidler}},
  \bibinfo {author} {\bibfnamefont {A.~V.}\ \bibnamefont {Allain}}, \bibinfo
  {author} {\bibfnamefont {D.~S.}\ \bibnamefont {Lembke}}, \bibinfo {author}
  {\bibfnamefont {A.}~\bibnamefont {Kis}},\ and\ \bibinfo {author}
  {\bibfnamefont {A.}~\bibnamefont {Imamo{\u{g}}lu}},\ }\bibfield  {title}
  {\enquote {\bibinfo {title} {{Optically active quantum dots in monolayer
  WSe2}},}\ }\href {https://doi.org/10.1038/nnano.2015.60} {\bibfield
  {journal} {\bibinfo  {journal} {Nature Nanotechnology}\ }\textbf {\bibinfo
  {volume} {10}},\ \bibinfo {pages} {491--496} (\bibinfo {year}
  {2015})}\BibitemShut {NoStop}%
\bibitem [{\citenamefont {He}\ \emph {et~al.}(2015)\citenamefont {He},
  \citenamefont {Clark}, \citenamefont {Schaibley}, \citenamefont {He},
  \citenamefont {Chen}, \citenamefont {Wei}, \citenamefont {Ding},
  \citenamefont {Zhang}, \citenamefont {Yao}, \citenamefont {Xu}, \citenamefont
  {Lu},\ and\ \citenamefont {Pan}}]{He2015}%
  \BibitemOpen
  \bibfield  {author} {\bibinfo {author} {\bibfnamefont {Y.-M.}\ \bibnamefont
  {He}}, \bibinfo {author} {\bibfnamefont {G.}~\bibnamefont {Clark}}, \bibinfo
  {author} {\bibfnamefont {J.~R.}\ \bibnamefont {Schaibley}}, \bibinfo {author}
  {\bibfnamefont {Y.}~\bibnamefont {He}}, \bibinfo {author} {\bibfnamefont
  {M.-C.}\ \bibnamefont {Chen}}, \bibinfo {author} {\bibfnamefont {Y.-J.}\
  \bibnamefont {Wei}}, \bibinfo {author} {\bibfnamefont {X.}~\bibnamefont
  {Ding}}, \bibinfo {author} {\bibfnamefont {Q.}~\bibnamefont {Zhang}},
  \bibinfo {author} {\bibfnamefont {W.}~\bibnamefont {Yao}}, \bibinfo {author}
  {\bibfnamefont {X.}~\bibnamefont {Xu}}, \bibinfo {author} {\bibfnamefont
  {C.-Y.}\ \bibnamefont {Lu}},\ and\ \bibinfo {author} {\bibfnamefont {J.-W.}\
  \bibnamefont {Pan}},\ }\bibfield  {title} {\enquote {\bibinfo {title}
  {{Single quantum emitters in monolayer semiconductors}},}\ }\href
  {https://doi.org/10.1038/nnano.2015.75} {\bibfield  {journal} {\bibinfo
  {journal} {Nature Nanotechnology}\ }\textbf {\bibinfo {volume} {10}},\
  \bibinfo {pages} {497--502} (\bibinfo {year} {2015})}\BibitemShut {NoStop}%
\bibitem [{\citenamefont {Roy}\ \emph {et~al.}(2014)\citenamefont {Roy},
  \citenamefont {Tosun}, \citenamefont {Kang}, \citenamefont {Sachid},
  \citenamefont {Desai}, \citenamefont {Hettick}, \citenamefont {Hu},\ and\
  \citenamefont {Javey}}]{Roy2014Field-EffectComponents}%
  \BibitemOpen
  \bibfield  {author} {\bibinfo {author} {\bibfnamefont {T.}~\bibnamefont
  {Roy}}, \bibinfo {author} {\bibfnamefont {M.}~\bibnamefont {Tosun}}, \bibinfo
  {author} {\bibfnamefont {J.~S.}\ \bibnamefont {Kang}}, \bibinfo {author}
  {\bibfnamefont {A.~B.}\ \bibnamefont {Sachid}}, \bibinfo {author}
  {\bibfnamefont {S.~B.}\ \bibnamefont {Desai}}, \bibinfo {author}
  {\bibfnamefont {M.}~\bibnamefont {Hettick}}, \bibinfo {author} {\bibfnamefont
  {C.~C.}\ \bibnamefont {Hu}},\ and\ \bibinfo {author} {\bibfnamefont
  {A.}~\bibnamefont {Javey}},\ }\bibfield  {title} {\enquote {\bibinfo {title}
  {{Field-Effect Transistors Built from All Two-Dimensional Material
  Components}},}\ }\href {https://doi.org/10.1021/nn501723y} {\bibfield
  {journal} {\bibinfo  {journal} {ACS Nano}\ }\textbf {\bibinfo {volume} {8}},\
  \bibinfo {pages} {6259--6264} (\bibinfo {year} {2014})}\BibitemShut {NoStop}%
\bibitem [{\citenamefont {Lee}\ \emph {et~al.}(2014)\citenamefont {Lee},
  \citenamefont {Lee}, \citenamefont {van~der Zande}, \citenamefont {Chen},
  \citenamefont {Li}, \citenamefont {Han}, \citenamefont {Cui}, \citenamefont
  {Arefe}, \citenamefont {Nuckolls}, \citenamefont {Heinz}, \citenamefont
  {Guo}, \citenamefont {Hone},\ and\ \citenamefont {Kim}}]{Lee2014}%
  \BibitemOpen
  \bibfield  {author} {\bibinfo {author} {\bibfnamefont {C.-H.}\ \bibnamefont
  {Lee}}, \bibinfo {author} {\bibfnamefont {G.-H.}\ \bibnamefont {Lee}},
  \bibinfo {author} {\bibfnamefont {A.~M.}\ \bibnamefont {van~der Zande}},
  \bibinfo {author} {\bibfnamefont {W.}~\bibnamefont {Chen}}, \bibinfo {author}
  {\bibfnamefont {Y.}~\bibnamefont {Li}}, \bibinfo {author} {\bibfnamefont
  {M.}~\bibnamefont {Han}}, \bibinfo {author} {\bibfnamefont {X.}~\bibnamefont
  {Cui}}, \bibinfo {author} {\bibfnamefont {G.}~\bibnamefont {Arefe}}, \bibinfo
  {author} {\bibfnamefont {C.}~\bibnamefont {Nuckolls}}, \bibinfo {author}
  {\bibfnamefont {T.~F.}\ \bibnamefont {Heinz}}, \bibinfo {author}
  {\bibfnamefont {J.}~\bibnamefont {Guo}}, \bibinfo {author} {\bibfnamefont
  {J.}~\bibnamefont {Hone}},\ and\ \bibinfo {author} {\bibfnamefont
  {P.}~\bibnamefont {Kim}},\ }\bibfield  {title} {\enquote {\bibinfo {title}
  {{Atomically thin p–n junctions with van der Waals heterointerfaces}},}\
  }\href {https://doi.org/10.1038/nnano.2014.150} {\bibfield  {journal}
  {\bibinfo  {journal} {Nature Nanotechnology}\ }\textbf {\bibinfo {volume}
  {9}},\ \bibinfo {pages} {676--681} (\bibinfo {year} {2014})}\BibitemShut
  {NoStop}%
\bibitem [{\citenamefont {Jariwala}\ \emph {et~al.}(2017)\citenamefont
  {Jariwala}, \citenamefont {Davoyan}, \citenamefont {Wong},\ and\
  \citenamefont {Atwater}}]{Jariwala2017VanOutlook}%
  \BibitemOpen
  \bibfield  {author} {\bibinfo {author} {\bibfnamefont {D.}~\bibnamefont
  {Jariwala}}, \bibinfo {author} {\bibfnamefont {A.~R.}\ \bibnamefont
  {Davoyan}}, \bibinfo {author} {\bibfnamefont {J.}~\bibnamefont {Wong}},\ and\
  \bibinfo {author} {\bibfnamefont {H.~A.}\ \bibnamefont {Atwater}},\
  }\bibfield  {title} {\enquote {\bibinfo {title} {{Van der Waals Materials for
  Atomically-Thin Photovoltaics: Promise and Outlook}},}\ }\href
  {https://doi.org/10.1021/acsphotonics.7b01103} {\bibfield  {journal}
  {\bibinfo  {journal} {ACS Photonics}\ }\textbf {\bibinfo {volume} {4}},\
  \bibinfo {pages} {2962--2970} (\bibinfo {year} {2017})}\BibitemShut {NoStop}%
\bibitem [{\citenamefont {Akama}\ \emph {et~al.}(2017)\citenamefont {Akama},
  \citenamefont {Okita}, \citenamefont {Nagai}, \citenamefont {Li},
  \citenamefont {Kaneko},\ and\ \citenamefont {Kato}}]{Akama2017}%
  \BibitemOpen
  \bibfield  {author} {\bibinfo {author} {\bibfnamefont {T.}~\bibnamefont
  {Akama}}, \bibinfo {author} {\bibfnamefont {W.}~\bibnamefont {Okita}},
  \bibinfo {author} {\bibfnamefont {R.}~\bibnamefont {Nagai}}, \bibinfo
  {author} {\bibfnamefont {C.}~\bibnamefont {Li}}, \bibinfo {author}
  {\bibfnamefont {T.}~\bibnamefont {Kaneko}},\ and\ \bibinfo {author}
  {\bibfnamefont {T.}~\bibnamefont {Kato}},\ }\bibfield  {title} {\enquote
  {\bibinfo {title} {{Schottky solar cell using few-layered transition metal
  dichalcogenides toward large-scale fabrication of semitransparent and
  flexible power generator}},}\ }\href
  {https://doi.org/10.1038/s41598-017-12287-6} {\bibfield  {journal} {\bibinfo
  {journal} {Scientific Reports}\ }\textbf {\bibinfo {volume} {7}},\ \bibinfo
  {pages} {11967} (\bibinfo {year} {2017})}\BibitemShut {NoStop}%
\bibitem [{\citenamefont {Liu}\ \emph {et~al.}(2014)\citenamefont {Liu},
  \citenamefont {Yan}, \citenamefont {Chen}, \citenamefont {Fan}, \citenamefont
  {Sun}, \citenamefont {Suh}, \citenamefont {Fu}, \citenamefont {Lee},
  \citenamefont {Zhou}, \citenamefont {Tongay}, \citenamefont {Ji},
  \citenamefont {Neaton},\ and\ \citenamefont {Wu}}]{Liu2014}%
  \BibitemOpen
  \bibfield  {author} {\bibinfo {author} {\bibfnamefont {K.}~\bibnamefont
  {Liu}}, \bibinfo {author} {\bibfnamefont {Q.}~\bibnamefont {Yan}}, \bibinfo
  {author} {\bibfnamefont {M.}~\bibnamefont {Chen}}, \bibinfo {author}
  {\bibfnamefont {W.}~\bibnamefont {Fan}}, \bibinfo {author} {\bibfnamefont
  {Y.}~\bibnamefont {Sun}}, \bibinfo {author} {\bibfnamefont {J.}~\bibnamefont
  {Suh}}, \bibinfo {author} {\bibfnamefont {D.}~\bibnamefont {Fu}}, \bibinfo
  {author} {\bibfnamefont {S.}~\bibnamefont {Lee}}, \bibinfo {author}
  {\bibfnamefont {J.}~\bibnamefont {Zhou}}, \bibinfo {author} {\bibfnamefont
  {S.}~\bibnamefont {Tongay}}, \bibinfo {author} {\bibfnamefont
  {J.}~\bibnamefont {Ji}}, \bibinfo {author} {\bibfnamefont {J.~B.}\
  \bibnamefont {Neaton}},\ and\ \bibinfo {author} {\bibfnamefont
  {J.}~\bibnamefont {Wu}},\ }\bibfield  {title} {\enquote {\bibinfo {title}
  {{Elastic Properties of Chemical-Vapor-Deposited Monolayer MoS <sub>2</sub> ,
  WS <sub>2</sub> , and Their Bilayer Heterostructures}},}\ }\href
  {https://doi.org/10.1021/nl501793a} {\bibfield  {journal} {\bibinfo
  {journal} {Nano Letters}\ }\textbf {\bibinfo {volume} {14}},\ \bibinfo
  {pages} {5097--5103} (\bibinfo {year} {2014})}\BibitemShut {NoStop}%
\bibitem [{\citenamefont {Lee}\ \emph {et~al.}(2008)\citenamefont {Lee},
  \citenamefont {Wei}, \citenamefont {Kysar}, \citenamefont {Hone},
  \citenamefont {Zettl}, \citenamefont {Guinea}, \citenamefont {Neto},\ and\
  \citenamefont {Crommie}}]{Lee2008}%
  \BibitemOpen
  \bibfield  {author} {\bibinfo {author} {\bibfnamefont {C.}~\bibnamefont
  {Lee}}, \bibinfo {author} {\bibfnamefont {X.}~\bibnamefont {Wei}}, \bibinfo
  {author} {\bibfnamefont {J.~W.}\ \bibnamefont {Kysar}}, \bibinfo {author}
  {\bibfnamefont {J.}~\bibnamefont {Hone}}, \bibinfo {author} {\bibfnamefont
  {A.}~\bibnamefont {Zettl}}, \bibinfo {author} {\bibfnamefont
  {F.}~\bibnamefont {Guinea}}, \bibinfo {author} {\bibfnamefont {A.~H.~C.}\
  \bibnamefont {Neto}},\ and\ \bibinfo {author} {\bibfnamefont {M.~F.}\
  \bibnamefont {Crommie}},\ }\bibfield  {title} {\enquote {\bibinfo {title}
  {{Measurement of the elastic properties and intrinsic strength of monolayer
  graphene.}}}\ }\href {https://doi.org/10.1126/science.1157996} {\bibfield
  {journal} {\bibinfo  {journal} {Science (New York, N.Y.)}\ }\textbf {\bibinfo
  {volume} {321}},\ \bibinfo {pages} {385--8} (\bibinfo {year}
  {2008})}\BibitemShut {NoStop}%
\bibitem [{\citenamefont {Rold{\'{a}}n}\ \emph {et~al.}(2015)\citenamefont
  {Rold{\'{a}}n}, \citenamefont {Castellanos-Gomez}, \citenamefont
  {Cappelluti},\ and\ \citenamefont {Guinea}}]{Roldan2015}%
  \BibitemOpen
  \bibfield  {author} {\bibinfo {author} {\bibfnamefont {R.}~\bibnamefont
  {Rold{\'{a}}n}}, \bibinfo {author} {\bibfnamefont {A.}~\bibnamefont
  {Castellanos-Gomez}}, \bibinfo {author} {\bibfnamefont {E.}~\bibnamefont
  {Cappelluti}},\ and\ \bibinfo {author} {\bibfnamefont {F.}~\bibnamefont
  {Guinea}},\ }\bibfield  {title} {\enquote {\bibinfo {title} {{Strain
  engineering in semiconducting two-dimensional crystals}},}\ }\href
  {https://doi.org/10.1088/0953-8984/27/31/313201} {\bibfield  {journal}
  {\bibinfo  {journal} {Journal of Physics: Condensed Matter}\ }\textbf
  {\bibinfo {volume} {27}},\ \bibinfo {pages} {313201} (\bibinfo {year}
  {2015})}\BibitemShut {NoStop}%
\bibitem [{\citenamefont {Feng}\ \emph {et~al.}(2012)\citenamefont {Feng},
  \citenamefont {Qian}, \citenamefont {Huang},\ and\ \citenamefont
  {Li}}]{Feng2012}%
  \BibitemOpen
  \bibfield  {author} {\bibinfo {author} {\bibfnamefont {J.}~\bibnamefont
  {Feng}}, \bibinfo {author} {\bibfnamefont {X.}~\bibnamefont {Qian}}, \bibinfo
  {author} {\bibfnamefont {C.-W.}\ \bibnamefont {Huang}},\ and\ \bibinfo
  {author} {\bibfnamefont {J.}~\bibnamefont {Li}},\ }\bibfield  {title}
  {\enquote {\bibinfo {title} {{Strain-engineered artificial atom as a
  broad-spectrum solar energy funnel}},}\ }\href
  {https://doi.org/10.1038/nphoton.2012.285} {\bibfield  {journal} {\bibinfo
  {journal} {Nature Photonics}\ }\textbf {\bibinfo {volume} {6}},\ \bibinfo
  {pages} {866--872} (\bibinfo {year} {2012})}\BibitemShut {NoStop}%
\bibitem [{\citenamefont {Conley}\ \emph {et~al.}(2013)\citenamefont {Conley},
  \citenamefont {Wang}, \citenamefont {Ziegler}, \citenamefont {Haglund},
  \citenamefont {Pantelides},\ and\ \citenamefont {Bolotin}}]{Conley2013}%
  \BibitemOpen
  \bibfield  {author} {\bibinfo {author} {\bibfnamefont {H.~J.}\ \bibnamefont
  {Conley}}, \bibinfo {author} {\bibfnamefont {B.}~\bibnamefont {Wang}},
  \bibinfo {author} {\bibfnamefont {J.~I.}\ \bibnamefont {Ziegler}}, \bibinfo
  {author} {\bibfnamefont {R.~F.}\ \bibnamefont {Haglund}}, \bibinfo {author}
  {\bibfnamefont {S.~T.}\ \bibnamefont {Pantelides}},\ and\ \bibinfo {author}
  {\bibfnamefont {K.~I.}\ \bibnamefont {Bolotin}},\ }\bibfield  {title}
  {\enquote {\bibinfo {title} {{Bandgap Engineering of Strained Monolayer and
  Bilayer MoS 2}},}\ }\href {https://doi.org/10.1021/nl4014748} {\bibfield
  {journal} {\bibinfo  {journal} {Nano Letters}\ }\textbf {\bibinfo {volume}
  {13}},\ \bibinfo {pages} {3626--3630} (\bibinfo {year} {2013})}\BibitemShut
  {NoStop}%
\bibitem [{\citenamefont {Niehues}\ \emph {et~al.}(2018)\citenamefont
  {Niehues}, \citenamefont {Schmidt}, \citenamefont {Dr{\"{u}}ppel},
  \citenamefont {Marauhn}, \citenamefont {Christiansen}, \citenamefont {Selig},
  \citenamefont {Bergh{\"{a}}user}, \citenamefont {Wigger}, \citenamefont
  {Schneider}, \citenamefont {Braasch}, \citenamefont {Koch}, \citenamefont
  {Castellanos-Gomez}, \citenamefont {Kuhn}, \citenamefont {Knorr},
  \citenamefont {Malic}, \citenamefont {Rohlfing}, \citenamefont {Michaelis~de
  Vasconcellos},\ and\ \citenamefont {Bratschitsch}}]{Niehues2018}%
  \BibitemOpen
  \bibfield  {author} {\bibinfo {author} {\bibfnamefont {I.}~\bibnamefont
  {Niehues}}, \bibinfo {author} {\bibfnamefont {R.}~\bibnamefont {Schmidt}},
  \bibinfo {author} {\bibfnamefont {M.}~\bibnamefont {Dr{\"{u}}ppel}}, \bibinfo
  {author} {\bibfnamefont {P.}~\bibnamefont {Marauhn}}, \bibinfo {author}
  {\bibfnamefont {D.}~\bibnamefont {Christiansen}}, \bibinfo {author}
  {\bibfnamefont {M.}~\bibnamefont {Selig}}, \bibinfo {author} {\bibfnamefont
  {G.}~\bibnamefont {Bergh{\"{a}}user}}, \bibinfo {author} {\bibfnamefont
  {D.}~\bibnamefont {Wigger}}, \bibinfo {author} {\bibfnamefont
  {R.}~\bibnamefont {Schneider}}, \bibinfo {author} {\bibfnamefont
  {L.}~\bibnamefont {Braasch}}, \bibinfo {author} {\bibfnamefont
  {R.}~\bibnamefont {Koch}}, \bibinfo {author} {\bibfnamefont {A.}~\bibnamefont
  {Castellanos-Gomez}}, \bibinfo {author} {\bibfnamefont {T.}~\bibnamefont
  {Kuhn}}, \bibinfo {author} {\bibfnamefont {A.}~\bibnamefont {Knorr}},
  \bibinfo {author} {\bibfnamefont {E.}~\bibnamefont {Malic}}, \bibinfo
  {author} {\bibfnamefont {M.}~\bibnamefont {Rohlfing}}, \bibinfo {author}
  {\bibfnamefont {S.}~\bibnamefont {Michaelis~de Vasconcellos}},\ and\ \bibinfo
  {author} {\bibfnamefont {R.}~\bibnamefont {Bratschitsch}},\ }\bibfield
  {title} {\enquote {\bibinfo {title} {{Strain Control of Exciton–Phonon
  Coupling in Atomically Thin Semiconductors}},}\ }\href
  {https://doi.org/10.1021/acs.nanolett.7b04868} {\bibfield  {journal}
  {\bibinfo  {journal} {Nano Letters}\ }\textbf {\bibinfo {volume} {18}},\
  \bibinfo {pages} {1751--1757} (\bibinfo {year} {2018})}\BibitemShut {NoStop}%
\bibitem [{\citenamefont {Christiansen}\ \emph {et~al.}(2017)\citenamefont
  {Christiansen}, \citenamefont {Selig}, \citenamefont {Bergh{\"{a}}user},
  \citenamefont {Schmidt}, \citenamefont {Niehues}, \citenamefont {Schneider},
  \citenamefont {Arora}, \citenamefont {de~Vasconcellos}, \citenamefont
  {Bratschitsch}, \citenamefont {Malic},\ and\ \citenamefont
  {Knorr}}]{Christiansen2017}%
  \BibitemOpen
  \bibfield  {author} {\bibinfo {author} {\bibfnamefont {D.}~\bibnamefont
  {Christiansen}}, \bibinfo {author} {\bibfnamefont {M.}~\bibnamefont {Selig}},
  \bibinfo {author} {\bibfnamefont {G.}~\bibnamefont {Bergh{\"{a}}user}},
  \bibinfo {author} {\bibfnamefont {R.}~\bibnamefont {Schmidt}}, \bibinfo
  {author} {\bibfnamefont {I.}~\bibnamefont {Niehues}}, \bibinfo {author}
  {\bibfnamefont {R.}~\bibnamefont {Schneider}}, \bibinfo {author}
  {\bibfnamefont {A.}~\bibnamefont {Arora}}, \bibinfo {author} {\bibfnamefont
  {S.~M.}\ \bibnamefont {de~Vasconcellos}}, \bibinfo {author} {\bibfnamefont
  {R.}~\bibnamefont {Bratschitsch}}, \bibinfo {author} {\bibfnamefont
  {E.}~\bibnamefont {Malic}},\ and\ \bibinfo {author} {\bibfnamefont
  {A.}~\bibnamefont {Knorr}},\ }\bibfield  {title} {\enquote {\bibinfo {title}
  {{Phonon Sidebands in Monolayer Transition Metal Dichalcogenides}},}\ }\href
  {https://doi.org/10.1103/PhysRevLett.119.187402} {\bibfield  {journal}
  {\bibinfo  {journal} {Physical Review Letters}\ }\textbf {\bibinfo {volume}
  {119}},\ \bibinfo {pages} {187402} (\bibinfo {year} {2017})}\BibitemShut
  {NoStop}%
\bibitem [{\citenamefont {Lloyd}\ \emph {et~al.}(2016)\citenamefont {Lloyd},
  \citenamefont {Liu}, \citenamefont {Christopher}, \citenamefont {Cantley},
  \citenamefont {Wadehra}, \citenamefont {Kim}, \citenamefont {Goldberg},
  \citenamefont {Swan},\ and\ \citenamefont {Bunch}}]{Lloyd2016}%
  \BibitemOpen
  \bibfield  {author} {\bibinfo {author} {\bibfnamefont {D.}~\bibnamefont
  {Lloyd}}, \bibinfo {author} {\bibfnamefont {X.}~\bibnamefont {Liu}}, \bibinfo
  {author} {\bibfnamefont {J.~W.}\ \bibnamefont {Christopher}}, \bibinfo
  {author} {\bibfnamefont {L.}~\bibnamefont {Cantley}}, \bibinfo {author}
  {\bibfnamefont {A.}~\bibnamefont {Wadehra}}, \bibinfo {author} {\bibfnamefont
  {B.~L.}\ \bibnamefont {Kim}}, \bibinfo {author} {\bibfnamefont {B.~B.}\
  \bibnamefont {Goldberg}}, \bibinfo {author} {\bibfnamefont {A.~K.}\
  \bibnamefont {Swan}},\ and\ \bibinfo {author} {\bibfnamefont {J.~S.}\
  \bibnamefont {Bunch}},\ }\bibfield  {title} {\enquote {\bibinfo {title}
  {{Band Gap Engineering with Ultralarge Biaxial Strains in Suspended Monolayer
  MoS 2}},}\ }\href {https://doi.org/10.1021/acs.nanolett.6b02615} {\bibfield
  {journal} {\bibinfo  {journal} {Nano Letters}\ }\textbf {\bibinfo {volume}
  {16}},\ \bibinfo {pages} {5836--5841} (\bibinfo {year} {2016})}\BibitemShut
  {NoStop}%
\bibitem [{\citenamefont {Lazi{\'{c}}}\ \emph {et~al.}(2014)\citenamefont
  {Lazi{\'{c}}}, \citenamefont {Violante}, \citenamefont {Cohen}, \citenamefont
  {Hey}, \citenamefont {Rapaport},\ and\ \citenamefont {Santos}}]{Lazic2014}%
  \BibitemOpen
  \bibfield  {author} {\bibinfo {author} {\bibfnamefont {S.}~\bibnamefont
  {Lazi{\'{c}}}}, \bibinfo {author} {\bibfnamefont {A.}~\bibnamefont
  {Violante}}, \bibinfo {author} {\bibfnamefont {K.}~\bibnamefont {Cohen}},
  \bibinfo {author} {\bibfnamefont {R.}~\bibnamefont {Hey}}, \bibinfo {author}
  {\bibfnamefont {R.}~\bibnamefont {Rapaport}},\ and\ \bibinfo {author}
  {\bibfnamefont {P.~V.}\ \bibnamefont {Santos}},\ }\bibfield  {title}
  {\enquote {\bibinfo {title} {{Scalable interconnections for remote indirect
  exciton systems based on acoustic transport}},}\ }\href
  {https://doi.org/10.1103/PhysRevB.89.085313} {\bibfield  {journal} {\bibinfo
  {journal} {Physical Review B}\ }\textbf {\bibinfo {volume} {89}},\ \bibinfo
  {pages} {085313} (\bibinfo {year} {2014})}\BibitemShut {NoStop}%
\bibitem [{\citenamefont {Manasevit}, \citenamefont {Gergis},\ and\
  \citenamefont {Jones}(1982)}]{Manasevit1982}%
  \BibitemOpen
  \bibfield  {author} {\bibinfo {author} {\bibfnamefont {H.~M.}\ \bibnamefont
  {Manasevit}}, \bibinfo {author} {\bibfnamefont {I.~S.}\ \bibnamefont
  {Gergis}},\ and\ \bibinfo {author} {\bibfnamefont {A.~B.}\ \bibnamefont
  {Jones}},\ }\bibfield  {title} {\enquote {\bibinfo {title} {{Electron
  mobility enhancement in epitaxial multilayer Si‐Si <sub> 1− <i>x</i>
  </sub> Ge <sub> <i>x</i> </sub> alloy films on (100) Si}},}\ }\href
  {https://doi.org/10.1063/1.93533} {\bibfield  {journal} {\bibinfo  {journal}
  {Applied Physics Letters}\ }\textbf {\bibinfo {volume} {41}},\ \bibinfo
  {pages} {464--466} (\bibinfo {year} {1982})}\BibitemShut {NoStop}%
\bibitem [{\citenamefont {People}\ \emph {et~al.}(1984)\citenamefont {People},
  \citenamefont {Bean}, \citenamefont {Lang}, \citenamefont {Sergent},
  \citenamefont {St{\"{o}}rmer}, \citenamefont {Wecht}, \citenamefont {Lynch},\
  and\ \citenamefont {Baldwin}}]{People1984ModulationHeterostructures}%
  \BibitemOpen
  \bibfield  {author} {\bibinfo {author} {\bibfnamefont {R.}~\bibnamefont
  {People}}, \bibinfo {author} {\bibfnamefont {J.~C.}\ \bibnamefont {Bean}},
  \bibinfo {author} {\bibfnamefont {D.~V.}\ \bibnamefont {Lang}}, \bibinfo
  {author} {\bibfnamefont {A.~M.}\ \bibnamefont {Sergent}}, \bibinfo {author}
  {\bibfnamefont {H.~L.}\ \bibnamefont {St{\"{o}}rmer}}, \bibinfo {author}
  {\bibfnamefont {K.~W.}\ \bibnamefont {Wecht}}, \bibinfo {author}
  {\bibfnamefont {R.~T.}\ \bibnamefont {Lynch}},\ and\ \bibinfo {author}
  {\bibfnamefont {K.}~\bibnamefont {Baldwin}},\ }\bibfield  {title} {\enquote
  {\bibinfo {title} {{Modulation doping in Ge <sub> <i>x</i> </sub> Si <sub>
  1− <i>x</i> </sub> /Si strained layer heterostructures}},}\ }\href
  {https://doi.org/10.1063/1.95074} {\bibfield  {journal} {\bibinfo  {journal}
  {Applied Physics Letters}\ }\textbf {\bibinfo {volume} {45}},\ \bibinfo
  {pages} {1231--1233} (\bibinfo {year} {1984})}\BibitemShut {NoStop}%
\bibitem [{\citenamefont {Castellanos-Gomez}\ \emph {et~al.}(2013)\citenamefont
  {Castellanos-Gomez}, \citenamefont {Rold{\'{a}}n}, \citenamefont
  {Cappelluti}, \citenamefont {Buscema}, \citenamefont {Guinea}, \citenamefont
  {van~der Zant},\ and\ \citenamefont {Steele}}]{Castellanos-Gomez2013}%
  \BibitemOpen
  \bibfield  {author} {\bibinfo {author} {\bibfnamefont {A.}~\bibnamefont
  {Castellanos-Gomez}}, \bibinfo {author} {\bibfnamefont {R.}~\bibnamefont
  {Rold{\'{a}}n}}, \bibinfo {author} {\bibfnamefont {E.}~\bibnamefont
  {Cappelluti}}, \bibinfo {author} {\bibfnamefont {M.}~\bibnamefont {Buscema}},
  \bibinfo {author} {\bibfnamefont {F.}~\bibnamefont {Guinea}}, \bibinfo
  {author} {\bibfnamefont {H.~S.~J.}\ \bibnamefont {van~der Zant}},\ and\
  \bibinfo {author} {\bibfnamefont {G.~A.}\ \bibnamefont {Steele}},\ }\bibfield
   {title} {\enquote {\bibinfo {title} {{Local Strain Engineering in Atomically
  Thin MoS 2}},}\ }\href {https://doi.org/10.1021/nl402875m} {\bibfield
  {journal} {\bibinfo  {journal} {Nano Letters}\ }\textbf {\bibinfo {volume}
  {13}},\ \bibinfo {pages} {5361--5366} (\bibinfo {year} {2013})}\BibitemShut
  {NoStop}%
\bibitem [{\citenamefont {Tyurnina}\ \emph {et~al.}(2019)\citenamefont
  {Tyurnina}, \citenamefont {Bandurin}, \citenamefont {Khestanova},
  \citenamefont {Kravets}, \citenamefont {Koperski}, \citenamefont {Guinea},
  \citenamefont {Grigorenko}, \citenamefont {Geim},\ and\ \citenamefont
  {Grigorieva}}]{Tyurnina2019}%
  \BibitemOpen
  \bibfield  {author} {\bibinfo {author} {\bibfnamefont {A.~V.}\ \bibnamefont
  {Tyurnina}}, \bibinfo {author} {\bibfnamefont {D.~A.}\ \bibnamefont
  {Bandurin}}, \bibinfo {author} {\bibfnamefont {E.}~\bibnamefont
  {Khestanova}}, \bibinfo {author} {\bibfnamefont {V.~G.}\ \bibnamefont
  {Kravets}}, \bibinfo {author} {\bibfnamefont {M.}~\bibnamefont {Koperski}},
  \bibinfo {author} {\bibfnamefont {F.}~\bibnamefont {Guinea}}, \bibinfo
  {author} {\bibfnamefont {A.~N.}\ \bibnamefont {Grigorenko}}, \bibinfo
  {author} {\bibfnamefont {A.~K.}\ \bibnamefont {Geim}},\ and\ \bibinfo
  {author} {\bibfnamefont {I.~V.}\ \bibnamefont {Grigorieva}},\ }\bibfield
  {title} {\enquote {\bibinfo {title} {{Strained Bubbles in van der Waals
  Heterostructures as Local Emitters of Photoluminescence with Adjustable
  Wavelength}},}\ }\href {https://doi.org/10.1021/acsphotonics.8b01497}
  {\bibfield  {journal} {\bibinfo  {journal} {ACS Photonics}\ }\textbf
  {\bibinfo {volume} {6}},\ \bibinfo {pages} {516--524} (\bibinfo {year}
  {2019})}\BibitemShut {NoStop}%
\bibitem [{\citenamefont {Castellanos-Gomez}\ \emph {et~al.}(2014)\citenamefont
  {Castellanos-Gomez}, \citenamefont {Buscema}, \citenamefont {Molenaar},
  \citenamefont {Singh}, \citenamefont {Janssen}, \citenamefont {van~der
  Zant},\ and\ \citenamefont
  {Steele}}]{Castellanos-Gomez2014DeterministicStamping}%
  \BibitemOpen
  \bibfield  {author} {\bibinfo {author} {\bibfnamefont {A.}~\bibnamefont
  {Castellanos-Gomez}}, \bibinfo {author} {\bibfnamefont {M.}~\bibnamefont
  {Buscema}}, \bibinfo {author} {\bibfnamefont {R.}~\bibnamefont {Molenaar}},
  \bibinfo {author} {\bibfnamefont {V.}~\bibnamefont {Singh}}, \bibinfo
  {author} {\bibfnamefont {L.}~\bibnamefont {Janssen}}, \bibinfo {author}
  {\bibfnamefont {H.~S.~J.}\ \bibnamefont {van~der Zant}},\ and\ \bibinfo
  {author} {\bibfnamefont {G.~A.}\ \bibnamefont {Steele}},\ }\bibfield  {title}
  {\enquote {\bibinfo {title} {{Deterministic transfer of two-dimensional
  materials by all-dry viscoelastic stamping}},}\ }\href
  {https://doi.org/10.1088/2053-1583/1/1/011002} {\bibfield  {journal}
  {\bibinfo  {journal} {2D Materials}\ }\textbf {\bibinfo {volume} {1}},\
  \bibinfo {pages} {011002} (\bibinfo {year} {2014})}\BibitemShut {NoStop}%
\bibitem [{See()}]{SeeOnline}%
  \BibitemOpen
  \bibfield  {title} {\enquote {\bibinfo {title} {{See Supplementary
  Information available online}},}\ }\href@noop {} {\bibinfo  {journal} {See
  Supplementary Information available online}\ }\BibitemShut {NoStop}%
\bibitem [{\citenamefont {Dukic}, \citenamefont {Adams},\ and\ \citenamefont
  {Fantner}(2015)}]{Dukic2015}%
  \BibitemOpen
\bibfield  {journal} {  }\bibfield  {author} {\bibinfo {author} {\bibfnamefont
  {M.}~\bibnamefont {Dukic}}, \bibinfo {author} {\bibfnamefont {J.~D.}\
  \bibnamefont {Adams}},\ and\ \bibinfo {author} {\bibfnamefont {G.~E.}\
  \bibnamefont {Fantner}},\ }\bibfield  {title} {\enquote {\bibinfo {title}
  {{Piezoresistive AFM cantilevers surpassing standard optical beam deflection
  in low noise topography imaging}},}\ }\href
  {https://doi.org/10.1038/srep16393} {\bibfield  {journal} {\bibinfo
  {journal} {Scientific Reports}\ }\textbf {\bibinfo {volume} {5}},\ \bibinfo
  {pages} {16393} (\bibinfo {year} {2015})}\BibitemShut {NoStop}%
\bibitem [{\citenamefont {Castellanos-Gomez}\ \emph {et~al.}(2012)\citenamefont
  {Castellanos-Gomez}, \citenamefont {Poot}, \citenamefont {Steele},
  \citenamefont {van~der Zant}, \citenamefont {Agra{\"{i}}t},\ and\
  \citenamefont {Rubio-Bollinger}}]{Castellanos-Gomez2012}%
  \BibitemOpen
  \bibfield  {author} {\bibinfo {author} {\bibfnamefont {A.}~\bibnamefont
  {Castellanos-Gomez}}, \bibinfo {author} {\bibfnamefont {M.}~\bibnamefont
  {Poot}}, \bibinfo {author} {\bibfnamefont {G.~A.}\ \bibnamefont {Steele}},
  \bibinfo {author} {\bibfnamefont {H.~S.~J.}\ \bibnamefont {van~der Zant}},
  \bibinfo {author} {\bibfnamefont {N.}~\bibnamefont {Agra{\"{i}}t}},\ and\
  \bibinfo {author} {\bibfnamefont {G.}~\bibnamefont {Rubio-Bollinger}},\
  }\bibfield  {title} {\enquote {\bibinfo {title} {{Elastic Properties of
  Freely Suspended MoS2 Nanosheets}},}\ }\href
  {https://doi.org/10.1002/adma.201103965} {\bibfield  {journal} {\bibinfo
  {journal} {Advanced Materials}\ }\textbf {\bibinfo {volume} {24}},\ \bibinfo
  {pages} {772--775} (\bibinfo {year} {2012})}\BibitemShut {NoStop}%
\bibitem [{\citenamefont {Zhang}, \citenamefont {Koutsos},\ and\ \citenamefont
  {Cheung}(2016)}]{Zhang2016}%
  \BibitemOpen
  \bibfield  {author} {\bibinfo {author} {\bibfnamefont {R.}~\bibnamefont
  {Zhang}}, \bibinfo {author} {\bibfnamefont {V.}~\bibnamefont {Koutsos}},\
  and\ \bibinfo {author} {\bibfnamefont {R.}~\bibnamefont {Cheung}},\
  }\bibfield  {title} {\enquote {\bibinfo {title} {{Elastic properties of
  suspended multilayer WSe2}},}\ }\href {https://doi.org/10.1063/1.4940982}
  {\bibfield  {journal} {\bibinfo  {journal} {Applied Physics Letters}\
  }\textbf {\bibinfo {volume} {108}},\ \bibinfo {pages} {042104} (\bibinfo
  {year} {2016})}\BibitemShut {NoStop}%
\bibitem [{\citenamefont {Vella}\ and\ \citenamefont
  {Davidovitch}(2017)}]{Vella2017}%
  \BibitemOpen
  \bibfield  {author} {\bibinfo {author} {\bibfnamefont {D.}~\bibnamefont
  {Vella}}\ and\ \bibinfo {author} {\bibfnamefont {B.}~\bibnamefont
  {Davidovitch}},\ }\bibfield  {title} {\enquote {\bibinfo {title}
  {{Indentation metrology of clamped, ultra-thin elastic sheets}},}\ }\href
  {https://doi.org/10.1039/c6sm02451c} {\bibfield  {journal} {\bibinfo
  {journal} {Soft Matter}\ }\textbf {\bibinfo {volume} {13}},\ \bibinfo {pages}
  {2264--2278} (\bibinfo {year} {2017})}\BibitemShut {NoStop}%
\bibitem [{\citenamefont {Kulig}\ \emph {et~al.}(2018)\citenamefont {Kulig},
  \citenamefont {Zipfel}, \citenamefont {Nagler}, \citenamefont {Blanter},
  \citenamefont {Sch{\"{u}}ller}, \citenamefont {Korn}, \citenamefont
  {Paradiso}, \citenamefont {Glazov},\ and\ \citenamefont
  {Chernikov}}]{Kulig2018}%
  \BibitemOpen
  \bibfield  {author} {\bibinfo {author} {\bibfnamefont {M.}~\bibnamefont
  {Kulig}}, \bibinfo {author} {\bibfnamefont {J.}~\bibnamefont {Zipfel}},
  \bibinfo {author} {\bibfnamefont {P.}~\bibnamefont {Nagler}}, \bibinfo
  {author} {\bibfnamefont {S.}~\bibnamefont {Blanter}}, \bibinfo {author}
  {\bibfnamefont {C.}~\bibnamefont {Sch{\"{u}}ller}}, \bibinfo {author}
  {\bibfnamefont {T.}~\bibnamefont {Korn}}, \bibinfo {author} {\bibfnamefont
  {N.}~\bibnamefont {Paradiso}}, \bibinfo {author} {\bibfnamefont {M.~M.}\
  \bibnamefont {Glazov}},\ and\ \bibinfo {author} {\bibfnamefont
  {A.}~\bibnamefont {Chernikov}},\ }\bibfield  {title} {\enquote {\bibinfo
  {title} {{Exciton Diffusion and Halo Effects in Monolayer Semiconductors}},}\
  }\href {https://doi.org/10.1103/PhysRevLett.120.207401} {\bibfield  {journal}
  {\bibinfo  {journal} {Physical Review Letters}\ }\textbf {\bibinfo {volume}
  {120}},\ \bibinfo {pages} {207401} (\bibinfo {year} {2018})}\BibitemShut
  {NoStop}%
\bibitem [{\citenamefont {Ovchinnikov}\ \emph {et~al.}(2014)\citenamefont
  {Ovchinnikov}, \citenamefont {Allain}, \citenamefont {Huang}, \citenamefont
  {Dumcenco},\ and\ \citenamefont {Kis}}]{Ovchinnikov2014Electricalsub2/sub}%
  \BibitemOpen
  \bibfield  {author} {\bibinfo {author} {\bibfnamefont {D.}~\bibnamefont
  {Ovchinnikov}}, \bibinfo {author} {\bibfnamefont {A.}~\bibnamefont {Allain}},
  \bibinfo {author} {\bibfnamefont {Y.-S.}\ \bibnamefont {Huang}}, \bibinfo
  {author} {\bibfnamefont {D.}~\bibnamefont {Dumcenco}},\ and\ \bibinfo
  {author} {\bibfnamefont {A.}~\bibnamefont {Kis}},\ }\bibfield  {title}
  {\enquote {\bibinfo {title} {{Electrical Transport Properties of Single-Layer
  WS <sub>2</sub>}},}\ }\href {https://doi.org/10.1021/nn502362b} {\bibfield
  {journal} {\bibinfo  {journal} {ACS Nano}\ }\textbf {\bibinfo {volume} {8}},\
  \bibinfo {pages} {8174--8181} (\bibinfo {year} {2014})}\BibitemShut {NoStop}%
\bibitem [{\citenamefont {Siviniant}\ \emph {et~al.}(1999)\citenamefont
  {Siviniant}, \citenamefont {Scalbert}, \citenamefont {Kavokin}, \citenamefont
  {Coquillat},\ and\ \citenamefont {Lascaray}}]{Siviniant1999ChemicalWells}%
  \BibitemOpen
  \bibfield  {author} {\bibinfo {author} {\bibfnamefont {J.}~\bibnamefont
  {Siviniant}}, \bibinfo {author} {\bibfnamefont {D.}~\bibnamefont {Scalbert}},
  \bibinfo {author} {\bibfnamefont {A.~V.}\ \bibnamefont {Kavokin}}, \bibinfo
  {author} {\bibfnamefont {D.}~\bibnamefont {Coquillat}},\ and\ \bibinfo
  {author} {\bibfnamefont {J.-P.}\ \bibnamefont {Lascaray}},\ }\bibfield
  {title} {\enquote {\bibinfo {title} {{Chemical equilibrium between excitons,
  electrons, and negatively charged excitons in semiconductor quantum
  wells}},}\ }\href {https://doi.org/10.1103/PhysRevB.59.1602} {\bibfield
  {journal} {\bibinfo  {journal} {Physical Review B}\ }\textbf {\bibinfo
  {volume} {59}},\ \bibinfo {pages} {1602--1604} (\bibinfo {year}
  {1999})}\BibitemShut {NoStop}%
\bibitem [{\citenamefont {Ross}\ \emph {et~al.}(2013)\citenamefont {Ross},
  \citenamefont {Wu}, \citenamefont {Yu}, \citenamefont {Ghimire},
  \citenamefont {Jones}, \citenamefont {Aivazian}, \citenamefont {Yan},
  \citenamefont {Mandrus}, \citenamefont {Xiao}, \citenamefont {Yao},\ and\
  \citenamefont {Xu}}]{Ross2013ElectricalSemiconductor}%
  \BibitemOpen
  \bibfield  {author} {\bibinfo {author} {\bibfnamefont {J.~S.}\ \bibnamefont
  {Ross}}, \bibinfo {author} {\bibfnamefont {S.}~\bibnamefont {Wu}}, \bibinfo
  {author} {\bibfnamefont {H.}~\bibnamefont {Yu}}, \bibinfo {author}
  {\bibfnamefont {N.~J.}\ \bibnamefont {Ghimire}}, \bibinfo {author}
  {\bibfnamefont {A.~M.}\ \bibnamefont {Jones}}, \bibinfo {author}
  {\bibfnamefont {G.}~\bibnamefont {Aivazian}}, \bibinfo {author}
  {\bibfnamefont {J.}~\bibnamefont {Yan}}, \bibinfo {author} {\bibfnamefont
  {D.~G.}\ \bibnamefont {Mandrus}}, \bibinfo {author} {\bibfnamefont
  {D.}~\bibnamefont {Xiao}}, \bibinfo {author} {\bibfnamefont {W.}~\bibnamefont
  {Yao}},\ and\ \bibinfo {author} {\bibfnamefont {X.}~\bibnamefont {Xu}},\
  }\bibfield  {title} {\enquote {\bibinfo {title} {{Electrical control of
  neutral and charged excitons in a monolayer semiconductor}},}\ }\href
  {https://doi.org/10.1038/ncomms2498} {\bibfield  {journal} {\bibinfo
  {journal} {Nature Communications}\ }\textbf {\bibinfo {volume} {4}},\
  \bibinfo {pages} {1474} (\bibinfo {year} {2013})}\BibitemShut {NoStop}%
\end{thebibliography}%

\end{document}